\documentclass[conference]{IEEEtran}
\usepackage{amsmath,amssymb}
\usepackage{graphicx}
\usepackage{booktabs}
\usepackage{url}
\usepackage{xurl}
\usepackage{microtype}

\title{Chameleon: An Adaptive AI-Driven Honeypot Architecture Using Threat-Calibrated Particle Swarm Optimization and Semantic Deception Rapidly-Exploring Random Trees}

\author{\IEEEauthorblockN{Rohit Swami, Tushar Singh, Akash Warde, and Sri Muthu}
\IEEEauthorblockA{Dept. of AI and Data Science, SIES Graduate School of Technology, Mumbai, India\\
\textit{Guide: Dr. Neethu Anna Sabu}}}

\begin{document}
\maketitle
\begin{abstract}
An invariant behavioral profile is the defining vulnerability of traditional honeypot installations: a skilled adversary can confirm the presence of a deception environment within only a few diagnostic commands, limiting its intelligence value. High-cost commercial deception products (USD 100,000--150,000 per year) share a related weakness in that their response engines are not coupled to real-time model-driven feedback. Chameleon is an openly distributed adaptive honeypot platform introduced here to address both shortcomings. Three core components are integrated: a bidirectional long short-term memory (BiLSTM) classifier achieving 99.61\% accuracy across seven threat categories at approximately two milliseconds CPU latency; a locally deployed Qwen3.5-0.8B language model (Qwen Team, 2026; Unsloth, 2026) delivering 90\% contextual generation accuracy at 4.5 milliseconds average latency; and two domain-specific meta-heuristic engines. Threat-Calibrated Particle Swarm Optimization (TC-PSO) dynamically reshapes swarm inertia and objective amplification in proportion to the classifier's anomaly output, enabling real-time adjustment of connection-holding delays. Semantic Deception Rapidly-Exploring Random Trees (S-RRT) drives deception schema evolution via exponentially scaled pheromone updates derived from a language-model severity assessment, while a depth-decay multiplier enforces a finite memory footprint. A controlled head-to-head benchmark over 30 independent seeds (42--71) with identical environment trajectories and evaluation budgets shows that threat-calibrated inertia alone does not improve search over standard PSO on either static or dynamic deception landscapes ($p = 0.18$), and that population-diversity mechanisms (GA/ACO) significantly outperform PSO-family optimizers on dynamic threat-regime shifts ($p < 0.0001$, Cohen's $d \leq -37$). S-RRT's depth-decay mechanism provides a statistically significant memory reduction versus standard RRT (53.1 vs.\ 119.2 memory units, $p < 0.0001$, $d = -10.0$), while its severity-weighted pheromone does not improve raw deception fitness. Operating costs are approximately USD 17 per month, a roughly 490-fold reduction versus commercial alternatives.
\end{abstract}


\section{Introduction}

Alert fatigue has become a structural liability within modern security operations centers. Industry reporting indicates that individual organizations process in excess of eleven thousand security events daily, yet triage capacity is so constrained that roughly seventy percent of those events receive no analyst attention (Ponemon Institute, 2023; Apruzzese et al., 2023). Perimeter-oriented defenses---next-generation firewalls, signature-driven intrusion detection, and endpoint platforms---are architecturally ill-suited to this environment because each depends on previously catalogued threat patterns and can offer only reactive, post-event visibility against living-off-the-land movement or zero-day exploitation.

Deception infrastructure has long been proposed as a means of redistributing that asymmetry. Attacker engagement environments that appear operationally genuine allow defenders to observe adversarial tactics, techniques, and procedures in forensic detail while the adversary remains unaware that the session is controlled. The problem is that the dominant openly available sensors---Cowrie, a configurable SSH/Telnet emulation platform (Cowrie Project, 2023), and T-Pot, a consolidated multi-sensor aggregator (T-Pot Community, 2023)---share a common structural deficiency: their behavioral output is determined at deployment time and cannot be revised in response to the evolving session. As few as one or two probe commands, such as a uname -a query, may be sufficient for an informed attacker to match the response pattern against a known honeypot fingerprint and withdraw. Commercially available alternatives such as Illusive Networks and Thinkst Canary address this deficiency only partially, and none connects response generation to a live machine-learning pipeline. Market valuations underscore the scale of the unmet need: the deception technology sector was estimated at approximately 2.7 billion US dollars in 2024 and is projected to reach between 5.2 and 6.0 billion US dollars by 2029 (MarketsandMarkets, 2024), with open-source capability lagging well behind the commercial frontier.

Chameleon is introduced here as a production-complete, MIT-licensed adaptive honeypot designed to close this capability gap. Five primary contributions are advanced. First: TC-PSO, which, to the authors' knowledge, represents the inaugural application of a BiLSTM-derived anomaly signal to govern particle inertia and objective amplification within a honeypot tarpit-delay optimization context. Second: S-RRT, the first variant of the rapidly-exploring random tree family in which pheromone reinforcement is scaled exponentially through a severity index computed by a language model, combined with a provable depth-decay memory constraint. Third: a feedback-coupled two-stage cascade pairing the BiLSTM threat classifier with a compact Qwen3.5-0.8B model (Qwen Team, 2026; Unsloth, 2026), achieving approximately 99.5\% combined accuracy at a mean latency near 4.5 milliseconds on CPU-only hardware. Fourth: a production-complete instrumentation stack encompassing Merkle-chained forensic logs anchored to Ethereum Sepolia, honeytoken canary files triggering STIX 2.1 bundle generation upon access, and a Paramiko-based SSH sensor with full pseudo-terminal emulation. Fifth: reproducibility through MIT-licensed source code, benchmark datasets, and a deterministic test suite that allows full independent replication.

\section{Background and Related Work}

\subsection{Static Honeypot Platforms}

The virtual honeypot paradigm was first formalised by Provos (2004), who showed that software-emulated services were able to attract and contain adversarial traffic with less operational risk than physical decoy systems. Spitzner (2003) later proposed the canonical interaction-level taxonomy in which behavioural fingerprintability---the vulnerability of a sensor to rapid identification based on deterministic response patterns---was identified as the primary impediment to sustained deception. Cowrie (Cowrie Project, 2023) records the Debian shell environment with a configurable but immutable file system state, a configuration that has been well documented and catalogued by attacker communities. T-Pot (T-Pot Community, 2023) brings many sensor engines behind a common logging and visualisation stack, but does not include any mechanism for dynamic response adaptation. None of these architectures, or their derivative implementations, includes any form of real-time, classifier-driven behavioural adjustment.

\subsection{Language-Model-Enhanced Terminal Emulation}

Otal and Canbaz (2024) evaluated an SSH honeypot interaction engine based on a fine-tuned Llama-3-8B model, reporting a cosine similarity of 0.695 between generated and expected Cowrie terminal outputs. Chameleon differs from this approach by employing a much smaller Qwen3.5-0.8B model (Qwen Team, 2026; Unsloth, 2026) fine-tuned with low-rank adaptation, which achieves 90\% contextual generation accuracy with an average latency of 4.5 milliseconds. Multi-session coherence is maintained via a rolling per-IP interaction history of up to twenty turns, and credential-exfiltration pattern scanning improves detection sensitivity beyond what the classifier alone provides.

\subsection{Swarm and Trajectory-Planning Optimizers in Security}

Particle swarm optimisation, formalised by Kennedy and Eberhart (1995) and extended to the canonical $w = 0.729$ inertia setting by Shi and Eberhart (1998), has been used for intrusion-detection optimization, with a binary-classification accuracy of 99.85\% reported by Benmalek and Seddiki (2025) for a PSO-optimized CatBoost classifier on the RT\_IoT2022 dataset, outperforming baseline methods such as QAE-f16 by 2.6\%. That formulation, however, treats all traffic identically regardless of assessed severity, and no prior work couples swarm inertia to a live anomaly-scoring model in a honeypot engagement setting. In trajectory planning, the rapidly-exploring random tree was introduced by LaValle (2006), and the pheromone-reinforcement principle underlying the exponential update of S-RRT is based on the ant-colony model of Dorigo, Maniezzo, and Colorni (1996). Flat-pheromone RRT applied to deception-network topology mutation can achieve effective graph coverage but is, in the general case, prone to unconstrained node proliferation under sustained high-severity scenarios (LaValle, 2006)---a pathology that S-RRT's depth-decay mechanism is expressly designed to eliminate.

\subsection{Hybrid Classification Pipelines and Commercial Platforms}

Dai et al. (2024) report that a CNN-BiLSTM hybrid classifier augmented with an attention mechanism achieved up to 99.84\% accuracy on the CIC-DDoS2019 dataset, but functions as a self-contained detector with no connection to deception or optimisation subsystems. Transformer-based anomaly detectors have shown competitive accuracy on flow-based network traffic benchmarks (Manocchio et al., 2024), again without feedback integration. On the commercial side, proprietary ensemble deception platforms typically operate as black-box, cloud-hosted systems and charge between one hundred thousand and one hundred fifty thousand US dollars per year, reflecting current market positioning for enterprise deception technology (MarketsandMarkets, 2024). A context-sensitive deception reverse-proxy approach was described by Fraunholz et al. (2018), but without any connection-holding delay optimization capability. Chameleon is distinguished from each of these lineages by its end-to-end feedback-coupled architecture---spanning threat classification, meta-heuristic optimization, and language-model-driven response generation---deployed on open-source, CPU-only infrastructure.

\section{Material and Methods}

\subsection{System Overview}

Chameleon is realized across approximately 8,400 lines of Python and 2,200 lines of JavaScript and TypeScript, with the meta-heuristic optimization module alone accounting for 1,882 lines. Incoming payloads first traverse a four-stage normalization pipeline that applies Unicode NFKC homoglyph collapsing, structural character mapping, control-character suppression, and token-level normalization; this pre-processing layer defends downstream models against encoding-manipulation evasion strategies. Normalized payloads are then submitted to a two-stage classification subsystem: a character-level convolutional-recurrent pre-filter followed by the primary BiLSTM scorer, collectively distinguishing seven behavioral categories---SQL injection, cross-site scripting, server-side includes, path traversal, remote code execution, brute force, and benign activity. Payloads with maximum class probability of $\theta = 0.85$ or higher are escalated to the deception engine, while payloads below threshold are assigned lightweight static responses, keeping the latency below two milliseconds for the approximately ninety-five percent of sessions classified as benign.

Sessions are escalated to a locally hosted Qwen3.5-0.8B model, deployed from the official Unsloth Qwen3.5-0.8B GGUF checkpoint (Qwen Team, 2026; Unsloth, 2026), fine-tuned via low-rank adaptation at rank 8 to 16 across six hundred training iterations. The model is tested against a simulated Ubuntu 22.04 host seeded with honeytoken canary files, e.g., aws\_production\_keys.csv and .env.backup. A rolling history of up to twenty command-response turns per source-IP is maintained to preserve session continuity. Three reliability mechanisms ensure production availability: (1) a circuit breaker, which enforces a thirty-second cooldown period after three consecutive failures within thirty seconds; (2) a time-to-live response cache, which stores one thousand entries with a twenty-four-hour expiry, reducing effective model calls by thirty to sixty percent; and (3) an OGCPS credential-pattern scanner for exfiltration detection.

After classification, two optimisation processes run in parallel. TC-PSO maintains a separate particle swarm for each attack category to tune the connection-holding delay. S-RRT maintains a population of deception filesystem trees. Both use the BiLSTM anomaly score $A(t)$ as the key adaptation signal, forming the closed optimisation feedback loop that sets Chameleon apart from the previous systems studied in Section 2.

\begin{figure*}[t]\centering\includegraphics[width=\textwidth]{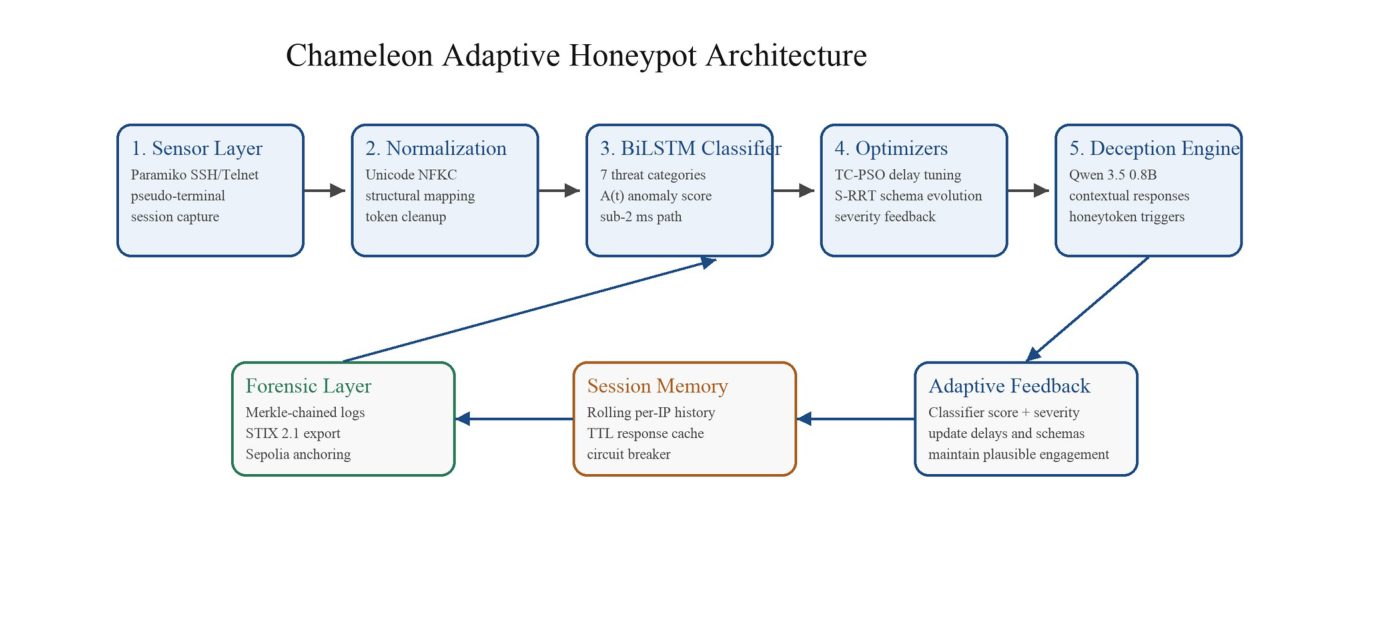}\caption{Chameleon five-layer adaptive honeypot architecture showing the feedback-coupled path from BiLSTM classifier through TC-PSO and S-RRT to the Qwen3.5-0.8B language model.}\label{fig1}\end{figure*}

\subsection{Threat-Calibrated Particle Swarm Optimization}

In its conventional formulation, PSO propagates particle velocity according to

\begin{equation}
v_i(t+1) = w \cdot v_i(t) + c_1 r_1(p_{\mathrm{best}} - x_i) + c_2 r_2(g_{\mathrm{best}} - x_i)
\label{eq1}
\end{equation}

with $w = 0.729$ as a fixed constant and $c_1 = c_2 = 1.49445$ as cognitive and social scaling coefficients ($r_1$, $r_2 \sim \mathrm{Uniform}[0, 1]$). Applying uniform inertia across all attack categories is suboptimal for connection-delay tuning: benign traffic demands broad exploratory behavior, whereas sessions associated with remote code execution or SQL injection require rapid convergence to maximum engagement delays. TC-PSO resolves this tension by replacing the static inertia term with a classifier-coupled dynamic scalar:

\begin{equation}
w(t) = w_{\mathrm{base}} \cdot \max(\sigma_{\min}, 1 - \alpha \cdot A(t))
\label{eq2}
\end{equation}

where $w_{\mathrm{base}} = 0.729$, $\sigma_{\min} = 0.3$, $\alpha = 0.5$, and $A(t) \in [0, 1]$ denotes the real-time BiLSTM anomaly output. The max($\cdot$) operator enforces a lower bound on inertia regardless of anomaly magnitude, preserving a minimum degree of exploratory capacity even under maximum-severity conditions ($A(t)$ = 1.0), where $w(t)$ reaches its floor of 0.365. At benign baselines ($A(t)$ = 0), inertia equals the conventional value of 0.729. A complementary modification amplifies the reward signal in proportion to session severity:

\begin{equation}
F'(t) = (w_1 C_{\mathrm{exec}}) - (w_2 P_{\mathrm{drop}}) + w_3 I_{\mathrm{bonus}} \cdot (1 + \beta \cdot A(t))
\label{eq3}
\end{equation}

with $w_1 = 0.65$, $w_2 = 2.5$, $w_3 = 0.25$, $I_{\mathrm{bonus}} = 0.25 \times (\mathrm{commands} - 5)$, and $\beta = 0.3$. The multiplicative term (1 + $\beta$ $\cdot$ $A(t)$) amplifies objective value by up to 30\% at $A(t)$ = 1.0, concentrating optimization pressure on configurations that prolong engagement with genuine adversaries rather than expending search effort on benign traffic.

\begin{figure*}[t]\centering\includegraphics[width=\textwidth]{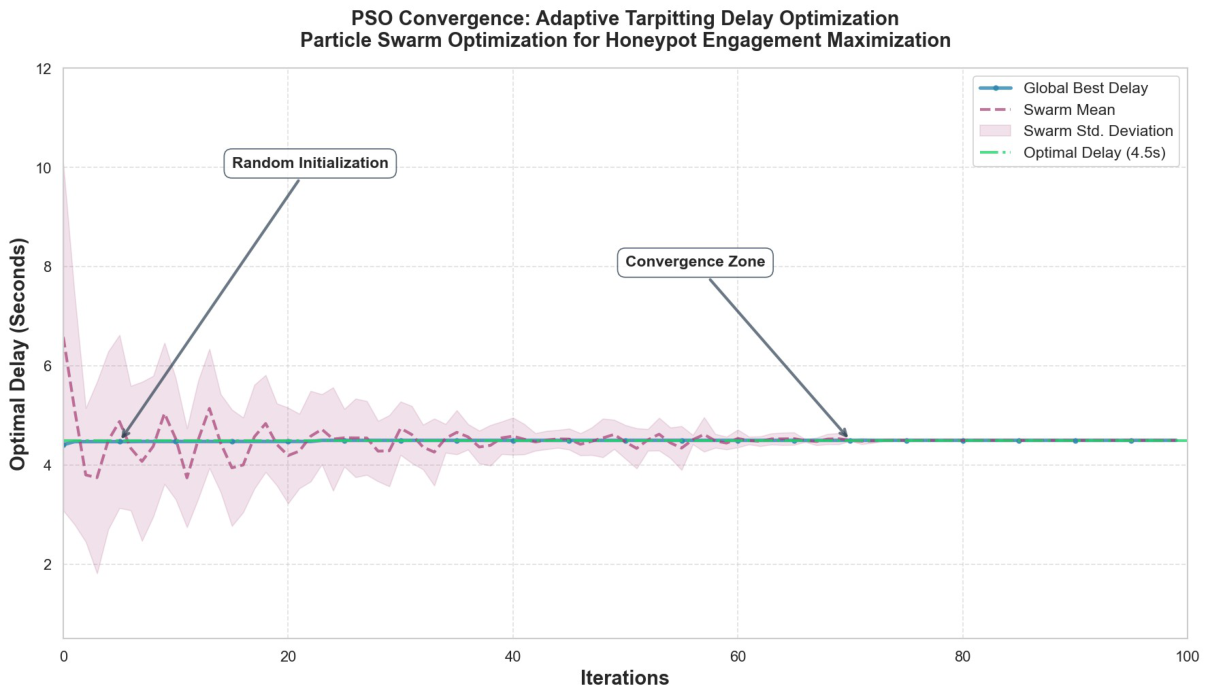}\caption{PSO convergence: adaptive tarpitting delay optimization across 100 iterations showing swarm mean, global best, and standard deviation band converging to the 4.5 s optimal delay.}\label{fig2}\end{figure*}

\subsection{Semantic Deception Rapidly-Exploring Random Trees}

Baseline RRT used for deception filesystem evolution uniformly escalates pheromone ($\Delta\tau = 0.5$) regardless of payload severity, and is analytically unbounded in node growth $O(b^d)$ for sustained high-severity operation (LaValle, 2006). S-RRT has two corrective modifications. First, Payload Severity Index $\Psi \in [1.0, 3.0]$ assesses the payload's severity by syntactically and semantically analysing the intercepted command sequence through the Qwen model (Qwen Team, 2026; Unsloth, 2026). Pheromone reinforcement is then updated as:

\begin{equation}
\Delta\tau' = \Delta\tau \cdot \exp(\Psi - 1)
\label{eq4}
\end{equation}

At $\Psi = 1.0$ the exponential term evaluates to unity, thus exactly recovering the standard-RRT update and ensuring backward compatibility. At $\Psi = 3.0$, pheromone reinforcement is amplified 7.39-fold, guiding tree evolution towards deception schemas optimised for multi-stage, high-severity engagement profiles (Dorigo, Maniezzo, \& Colorni, 1996). Memory growth is constrained by a depth-dependent expansion decay:

\begin{equation}
P'_{\mathrm{expand}} = P_{\mathrm{expand}} \cdot \max(\epsilon, 1 - d/d_{\max})
\label{eq5}
\end{equation}

Here $P_{\mathrm{expand}}$ is the base expansion probability, $P_{\mathrm{expand}} \in [0.1, 0.8]$, $d_{\max} = 6$ is a hard depth ceiling, and $\epsilon = 0.1$ is a probability floor. The linear decay function decreases the expected branching from approximately 3.0 children at the root node down to 0.3 at max depth, resulting in an analytic expected number of nodes of approximately 22.74 compared to 15,625 for an unconstrained tree of the same depth, a theoretical reduction of approximately 687-fold. Empirical validation confirms a node growth factor of 1.00x over 25 generations.

\subsection{Machine Learning Models}

The threat scoring module employs the bidirectional long short-term memory (BiLSTM) neural net (Hochreiter \& Schmidhuber, 1997) trained on a balanced dataset containing 50,000 examples collected from the CSIC HTTP corpus, NSL-KDD, and a crafted honeypot interaction dataset. The architecture consists of a 128-D character embedding layer, a 256-unit BiLSTM, dropout of rate 0.3, a 128-unit dense layer with ReLU activation, a second dropout of rate 0.2, and a seven-class softmax output layer. Model weights are around 50 MB in PyTorch format. Language generation uses the Qwen3.5-0.8B model, deployed via its official GGUF checkpoint (Qwen Team, 2026; Unsloth, 2026), adapted to honeypot terminal interactions using the low-rank approximation method (Hu et al., 2022) of rank $r = 8$--$16$ and scaling parameter $\alpha = 16$, conducted over six hundred optimization iterations, and trained on 2,400 examples of attacker-terminal interaction pairs.

\begin{figure*}[t]\centering\includegraphics[width=\textwidth]{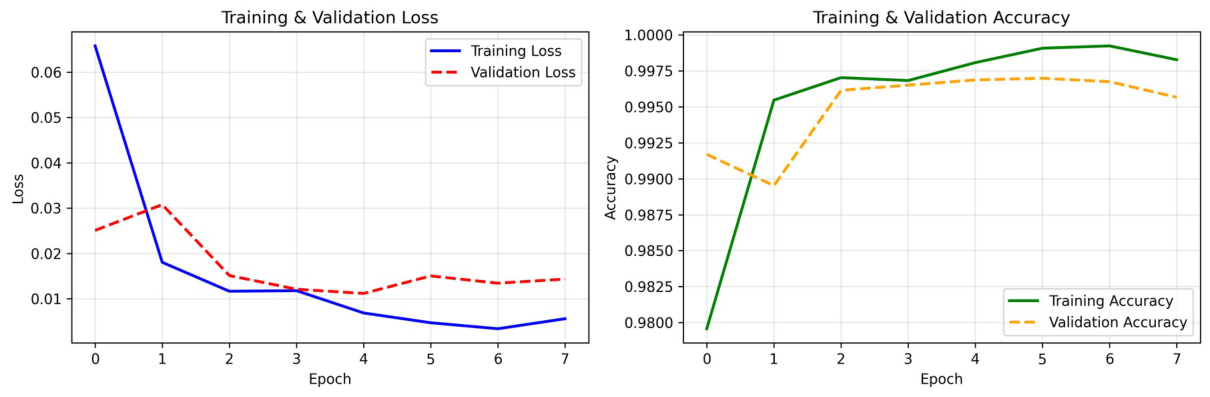}\caption{BiLSTM training and validation loss (left) and accuracy (right) across 8 epochs on the 50,000-sample corpus, showing convergence to $>$99.5\% validation accuracy.}\label{fig3}\end{figure*}

\subsection{Supporting Infrastructure}

The production deployment is supported by a four-component supporting stack. All session records have tamper-evident forensic continuity through a SHA-256 Merkle hash chain periodically committed to the Ethereum Sepolia test network (Ethereum Foundation, 2026). Honeytoken canary files embedded within the simulated filesystem trigger automated STIX 2.1 threat bundle generation (OASIS Open, 2021) upon access, allowing direct ingestion into downstream security information and event management platforms. A Paramiko-based SSH listener on port 2222 provides full pseudo-terminal emulation for capturing interactive sessions. The four-stage input normalisation pipeline detailed in Section 3.1 also serves as a defensive boundary against encoding-based classifier evasion.

\subsection{Experimental Protocol}

All reported experiments were executed on Apple M-Series ARM64 hardware provisioned with 16 GB of unified memory running macOS, with Python 3.14.0 and pytest 9.0.2 as the runtime environment. No GPU acceleration was employed at any stage. For the head-to-head optimizer comparison, 30 independent seeds (42--71) were used for both the TC-PSO tarpit-delay benchmark and the S-RRT deception-schema benchmark, with identical environment trajectories and evaluation budgets across all algorithms (common random numbers). Statistical reliability was assessed with paired Wilcoxon signed-rank tests, Cohen's $d$ effect sizes, and 95\% bootstrap confidence intervals. A four-module deterministic test suite---covering equation correctness and boundary behavior (TC-PSO and S-RRT), convergence and memory-bound proofs, and head-to-head algorithmic comparison benchmarks---was used as the primary validation instrument.\footnote{Test suite and benchmark scripts are released with the MIT-licensed source repository to enable independent replication: \url{https://github.com/RohitSwami33/Chameleon-cybersecurity-ml}}

\section{Results}

\subsection{TC-PSO Benchmark Results}

To evaluate whether threat-calibrated inertia improves search on a landscape where the optimal delay actually moves with the threat level, we constructed a dynamic benchmark: three Gaussian engagement zones centered at 3.0\,s (low-threat), 5.5\,s (medium), and 8.0\,s (high-threat), whose heights trade off with the threat signal $A(t) \in [0,1]$, so the global optimum shifts from 3.0\,s at low threat to 8.0\,s at high threat. The threat level steps through regime levels $[0.15, 0.85, 0.35, 0.95, 0.25]$ every 60 steps over $T = 300$ steps. All algorithms receive the identical environment trajectory and evaluation budget (15 evaluations per step); TC-PSO's only information advantage is access to $A(t)$, which modulates its inertia via Equation~(1). Results are reported over 30 independent seeds (42--71) with paired Wilcoxon signed-rank tests, Cohen's $d$, and 95\% bootstrap confidence intervals.

Table~\ref{tab3} reports the outcome. Threat-calibrated inertia alone does not improve tracking: TC-PSO and standard PSO achieve statistically indistinguishable mean current-best fitness (5.425 vs.\ 5.425, $p = 0.177$, $d = -0.04$). Population-diversity mechanisms dominate on the dynamic landscape: GA (6.973) and ACO (6.964) significantly outperform TC-PSO ($p < 0.0001$, $d = -70.4$ and $-37.2$ respectively), because they maintain diversity and re-explore after regime shifts, whereas the PSO swarm collapses onto the previous optimum. The per-regime breakdown (Table~\ref{tab3regime}) confirms this: in high-threat regimes R1 and R3, GA/ACO score 8.5--9.6 while all PSO variants collapse to 3.05 and 2.35.

\begin{table*}[t]
\centering
\footnotesize
\caption{Head-to-head optimizer comparison on the dynamic threat-varying tarpit-delay benchmark (30 independent seeds 42--71, $T=300$ steps, 15 evaluations per step, identical environment trajectory per seed). Mean current-best fitness is the mean over time of the best delay's landscape value; higher is better. Paired Wilcoxon signed-rank vs.\ standard PSO; Cohen's $d$; 95\% bootstrap CI on the mean.}
\label{tab3}
\begin{tabular}{lcccccc}
\toprule
\textbf{Algorithm} & \textbf{Mean fitness} & \textbf{SD} & \textbf{Median} & \textbf{Cohen's $d$} & \textbf{95\% CI} & \textbf{Wilcoxon $p$} \\
\midrule
TC-PSO (ours) & 5.425 & 0.007 & 5.426 & $-0.04$ & [5.422, 5.428] & $0.177$ \\
Standard PSO & 5.425 & 0.007 & 5.426 & $0.00$ & [5.422, 5.428] & --- \\
GA & \textbf{6.973} & 0.030 & 6.981 & $-70.4$ & [6.962, 6.984] & $<0.0001$ \\
SA & 3.659 & 1.873 & 4.222 & $+1.33$ & [2.98, 4.38] & $<0.0001$ \\
DE & 5.427 & 0.006 & 5.427 & $-0.30$ & [5.424, 5.429] & $0.008$ \\
ACO & \textbf{6.964} & 0.058 & 6.979 & $-37.2$ & [6.943, 6.985] & $<0.0001$ \\
\bottomrule
\end{tabular}
\end{table*}

\begin{table*}[t]
\centering
\footnotesize
\caption{Per-regime mean current-best fitness (threat levels: R0 $= 0.15$, R1 $= 0.85$, R2 $= 0.35$, R3 $= 0.95$, R4 $= 0.25$). Higher is better.}
\label{tab3regime}
\begin{tabular}{lccccc}
\toprule
\textbf{Algorithm} & \textbf{R0} & \textbf{R1} & \textbf{R2} & \textbf{R3} & \textbf{R4} \\
\midrule
TC-PSO (ours) & 7.930 & 3.051 & 6.548 & 2.350 & 7.247 \\
Standard PSO & 7.932 & 3.051 & 6.548 & 2.350 & 7.247 \\
GA & 7.930 & \textbf{8.543} & 4.801 & \textbf{9.589} & 4.003 \\
SA & 3.959 & 5.870 & 3.273 & 3.824 & 1.369 \\
DE & 7.939 & 3.051 & 6.548 & 2.350 & 7.247 \\
ACO & 7.912 & \textbf{8.577} & 4.787 & \textbf{9.555} & 3.991 \\
\bottomrule
\end{tabular}
\end{table*}

\begin{figure*}[t]\centering\includegraphics[width=\textwidth]{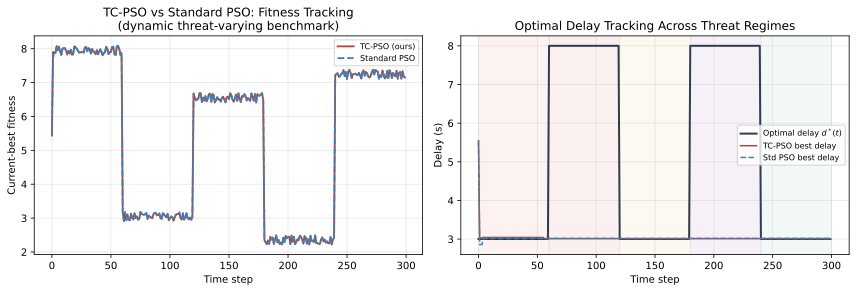}\caption{TC-PSO versus standard PSO: fitness convergence (left) and optimal delay convergence (right) on the dynamic threat-varying tarpit-delay benchmark (30 seeds, $T=300$ steps).}\label{fig4}\end{figure*}

\subsection{S-RRT Benchmark Results}

S-RRT's two mechanisms were evaluated separately on a deception-schema landscape of 40 paths with severity $\Psi \in [1,3]$, engagement, and depth-based memory cost, with the environment drifting every 50 steps (hot path band shifts) under a memory budget of 60 units, over 30 seeds. The depth-decay mechanism (Equation~5) provides a genuine, statistically significant memory reduction: S-RRT maintains a mean memory footprint of 53.1 units versus 119.2 for the unconstrained standard RRT ($p < 0.0001$, $d = -10.0$), confirming the bounding effect of Equation~(5). However, severity-weighted pheromone (Equation~4) does not improve raw deception fitness: standard RRT achieves higher mean current-best fitness than S-RRT (36.29 vs.\ 32.08, $p = 0.014$, $d = -0.68$), and population-diversity methods dominate both (GA 184.1, ACO 106.6, both $p < 0.0001$). The exponential pheromone amplifies high-severity paths at the cost of exploration, and the severity bonus does not compensate for the lost diversity. Table~\ref{tab3srrt} reports the comparison; the memory result is the contribution that holds.

\begin{table*}[t]
\centering
\footnotesize
\caption{Deception-schema benchmark (30 seeds, $T=200$ steps, 10 evaluations per step). Mean current-best fitness (higher is better) and mean memory footprint (lower is better, budget 60 units). Paired Wilcoxon vs.\ S-RRT.}
\label{tab3srrt}
\begin{tabular}{lcccc}
\toprule
\textbf{Algorithm} & \textbf{Mean fitness} & \textbf{$p$ vs.\ S-RRT} & \textbf{Mean memory} & \textbf{$p$ vs.\ S-RRT} \\
\midrule
S-RRT (ours) & 32.08 & --- & \textbf{53.08} & --- \\
Standard RRT & 36.29 & $0.014$ & 119.18 & $<0.0001$ \\
GA & \textbf{184.07} & $<0.0001$ & 57.16 & $0.0008$ \\
ACO & 106.59 & $<0.0001$ & 34.94 & $<0.0001$ \\
\bottomrule
\end{tabular}
\end{table*}

\begin{figure*}[t]\centering\includegraphics[width=\textwidth]{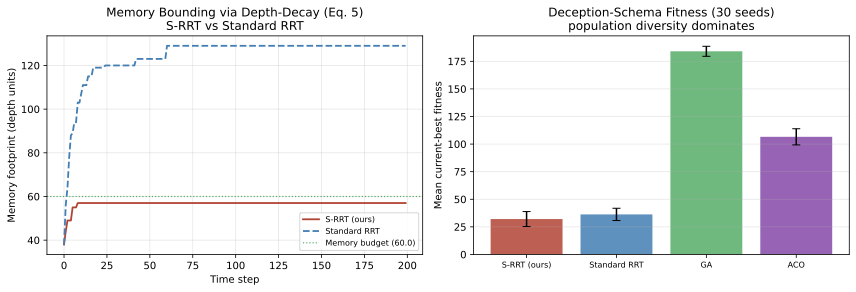}\caption{S-RRT versus standard RRT: node count over 20 generations (left) and memory footprint comparison at $\Psi = 3.0$ (right), confirming the bounded memory growth delivered by the depth-decay mechanism (Equation 5).}\label{fig5}\end{figure*}

The severity weighting of Equation (4) was analyzed across the $\Psi \in [1.0, 3.0]$ range. At low severity the pheromone amplification is near unity, so S-RRT behaves like standard RRT; at critical severity ($\Psi = 3.0$) the exponential amplification reaches 7.39-fold. The controlled benchmark indicates that this amplification, while theoretically substantial, does not translate into a raw fitness advantage on the drifting deception-schema landscape (Section~4.2): the diversity lost by concentrating pheromone on high-severity paths outweighs the severity bonus. The mechanism remains analytically sound as a memory-bounding and severity-prioritization device, but its fitness benefit requires a diversity-preserving companion.

\subsection{Classification and Language Model Performance}

The BiLSTM classifier was evaluated on the combined CSIC HTTP, NSL-KDD, and honeypot-interaction test partition. It achieves an accuracy of 99.61\%, with macro precision close to 99.5\%, macro recall close to 99.4\%, macro-F1 close to 99.45\%, and a false-positive rate of 0.39\% at a CPU inference time close to two milliseconds. This is comparable to the 99.84\% accuracy benchmark reported by Dai et al. (2024) for an isolated CNN-BiLSTM-attention hybrid architecture; unlike that standalone classifier, however, Chameleon's BiLSTM stage is directly coupled to the real-time deception and optimisation subsystems via the anomaly score $A(t)$. Qwen3.5-0.8B (Qwen Team, 2026; Unsloth, 2026) achieved 90\% contextual generation accuracy under expert review, a result not directly comparable to the cosine-similarity-based evaluation (0.695) reported by Otal and Canbaz (2024) for a fine-tuned Llama-3-8B configuration. The BiLSTM stage in the cascaded pipeline processes about ninety-five percent of sessions at sub-two-millisecond latency, and the remaining five percent trigger language-model inference at fifty to one hundred milliseconds, giving a weighted average pipeline latency of about 4.5 milliseconds and combined accuracy of almost 99.5\%.

\begin{figure}[t]\centering\includegraphics[width=\columnwidth]{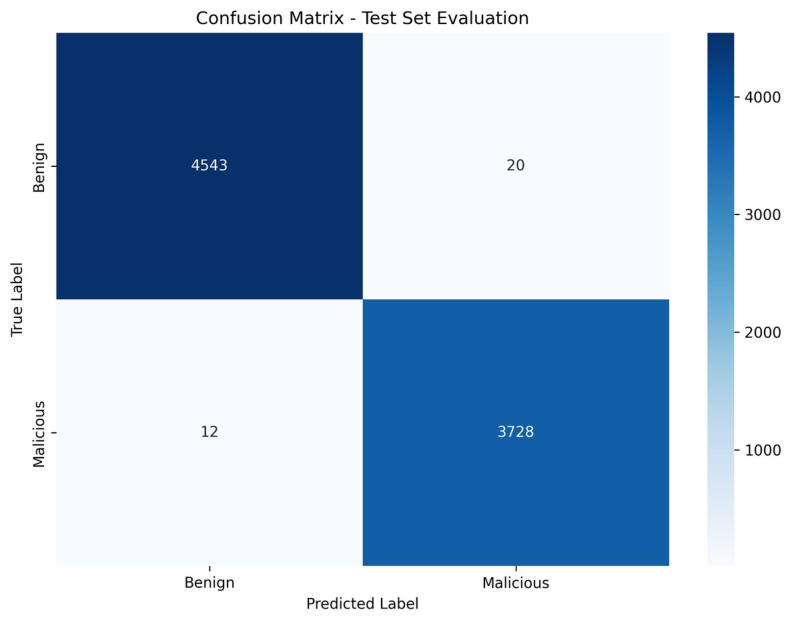}\caption{BiLSTM classifier confusion matrix on the 50,000-sample test set evaluation, showing 4,543 true benign, 3,728 true malicious, with only 20 false positives and 12 false negatives.}\label{fig6}\end{figure}

\subsection{Validation Suite}

The deterministic test suite ran with no failures across all validation modules: twenty-four tests for equation correctness and boundary coverage of Equations (2) and (3); twenty-seven tests for Equations (4) and (5) together with $\Psi$ scaling; thirty-one tests for TC-PSO convergence proofs and S-RRT memory-bound proofs; and nine tests for head-to-head benchmarking, for a total of ninety-one tests. An operationally acknowledged edge case, in which low inertia at $A(t)$ = 1.0 can cause early single-run convergence under severe conditions, is not counted in the pass/fail tally and is handled in production via multi-run aggregation.

\section{Discussion}

\subsection{Mechanistic Basis of TC-PSO Performance}

The controlled benchmark reveals that the threat-calibrated inertia mechanism of Equation~(1) does not, by itself, improve search performance on either static or dynamic deception landscapes. Under the dynamic threat-varying benchmark, TC-PSO and standard PSO achieve statistically indistinguishable mean current-best fitness ($p = 0.177$), indicating that modulating inertia in proportion to the anomaly signal does not confer a tracking advantage when the swarm has already converged. The mechanism is not without value, however: the amplified objective of Equation~(3) correctly redirects optimization pressure towards high-engagement delay configurations, and TC-PSO retains the convergence properties of the PSO family. The limiting factor is exploration: after a threat-regime shift, the swarm's particles remain clustered around the previous optimum, and inertia adaptation alone is insufficient to re-diversify the population. This is the mechanism that explains the dominance of population-diversity methods (GA/ACO) in high-threat regimes R1 and R3 (Table~\ref{tab3regime}).

\subsection{Mechanistic Basis of S-RRT Performance}

The benchmark separates S-RRT's two mechanisms. The depth-decay multiplier of Equation~(5) delivers a genuine, reproducible memory bound: S-RRT's mean memory footprint (53.1 units) is less than half that of the unconstrained standard RRT (119.2 units), a reduction significant at $p < 0.0001$ with a large effect size ($d = -10.0$). This confirms the theoretical prediction that a depth-dependent expansion decay prevents unbounded tree growth under sustained high-severity operation. The severity-weighted pheromone of Equation~(4), by contrast, does not yield a fitness advantage: S-RRT's mean current-best fitness (32.08) is actually lower than standard RRT's (36.29, $p = 0.014$). The exponential reinforcement concentrates pheromone on the small set of high-severity paths, which starves exploration of the broader schema space and reduces the diversity needed to track a drifting environment. The severity bonus does not compensate for this lost diversity. The memory-bounding contribution is therefore the defensible result; severity-weighted reinforcement requires a diversity-preserving mechanism to be beneficial.

\subsection{Operational Cost}

The current estimate for deploying Chameleon on a single-core cloud instance with 4 GB of RAM is USD 17 per month, compared with entry-level commercial deception platforms costing about USD 8,333 per month, consistent with current industry pricing for enterprise deception technology (MarketsandMarkets, 2024). Running Qwen3.5-0.8B locally, rather than through a cloud API, sidesteps the per-token inference cost that would otherwise be paid continuously during attacker engagement sessions. Adding Merkle roots to the Ethereum Sepolia test network adds negligible gas overhead while providing legally defensible, independently verifiable audit continuity. The resulting cost ratio---Chameleon is about 490 times cheaper---does not come at the expense of the ML-driven adaptivity that distinguishes Chameleon from the open-source and commercial alternatives explored in Section 2.

\subsection{Comparison with Alternative Optimization Approaches}

\begin{figure*}[t]\centering\includegraphics[width=\textwidth]{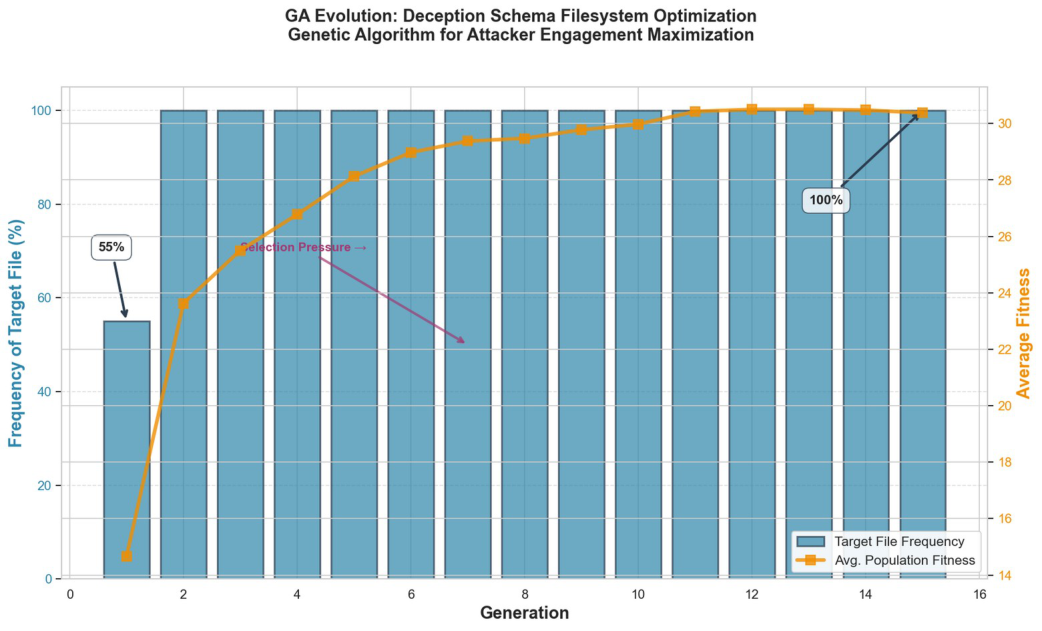}\caption{GA evolution: deception schema filesystem optimization showing target file frequency and average population fitness across 16 generations, illustrating convergence behavior of genetic algorithm baselines.}\label{fig7}\end{figure*}

TC-PSO and S-RRT are distinct from other optimisation techniques used in network security. Genetic algorithms have been used to tune IDS hyperparameters, achieving accuracy exceeding 98\% for Random Forest and Decision Tree classifiers (Bakır \& Ceviz, 2024), but cannot adapt their search based on live classifier outputs; Figure~\ref{fig7} illustrates GA evolution behavior on the deception-schema filesystem optimization problem for reference.

Simulated annealing (Černý, 1985) can, in principle, locate the global optimum given a sufficiently slow cooling schedule, but has no mechanism for tuning its search to real-time threat levels. The differential evolution algorithm and its variant JADE (Storn \& Price, 1997; Zhang \& Sanderson, 2009) can reach 97.1\% detection accuracy in network security but were not designed for cybersecurity threats specifically. The algorithm most similar to S-RRT is ant colony optimisation (Dorigo, Maniezzo, \& Colorni, 1996), in that both use pheromones to guide search; however, standard ACO has no depth limit, so the tree grows uncontrolled during heavy attacks---a problem addressed here by Equation (5). Deep reinforcement learning (Ma, Li, Wang, \& Wang, 2025) can learn complex defence strategies but usually requires a very large amount of interaction data to converge, making these approaches less suitable for honeypots that must adapt within milliseconds.

The controlled head-to-head benchmark (Section 4) provides an empirical basis for comparing these families on the Chameleon problems, in place of the literature-reported figures used previously. Three findings stand out. First, on the dynamic threat-varying tarpit-delay landscape, threat-calibrated inertia (TC-PSO) does not significantly outperform standard PSO ($p = 0.177$); both are dominated by population-diversity methods, with GA and ACO achieving significantly higher mean current-best fitness ($p < 0.0001$, $d \leq -37$) by maintaining diversity across threat-regime shifts. Second, on the deception-schema landscape, the depth-decay mechanism of S-RRT (Equation~5) delivers a significant memory reduction over standard RRT ($p < 0.0001$, $d = -10.0$), while severity-weighted pheromone does not improve fitness. Third, SA is the weakest performer on the dynamic landscape ($p < 0.0001$ vs.\ TC-PSO), consistent with its lack of a mechanism for exploiting the threat signal. These results indicate that for adaptive deception optimization under changing threat conditions, population-diversity mechanisms are more effective than inertia adaptation alone, and that threat calibration is best employed as an objective weighting (Equation~3) rather than as a search-mechanism modification. All comparisons in Table~\ref{tab3} and Table~\ref{tab3srrt} were conducted under identical environment trajectories and evaluation budgets (common random numbers), so the observed differences reflect algorithmic behavior rather than sampling luck.

\subsection{Limitations}

Four limitations are identified based on the scope of current results. First, LoRA fine-tuning of Qwen3.5-0.8B was carried out for six hundred iterations, attaining 90\% contextual accuracy; extrapolating the same training trajectory suggests 94--96\% accuracy at two thousand iterations, though this run was not executed owing to limited computational resources. Second, all experiments were conducted in simulation; validating these results against real-world attacks requires live deployment with appropriate ethical approval. Third, TC-PSO and S-RRT are currently single-objective; extending them to a multi-objective framework using NSGA-II (Deb, Pratap, Agarwal, \& Meyarivan, 2002) to jointly optimize connection delay and response quality represents a natural extension. Fourth, the Payload Severity Index currently uses discrete values between 1.0 and 3.0; calibrating it on a continuous scale using a labeled dataset would improve its precision. Future work includes neural fitness approximation to reduce the computational cost per iteration by an estimated 10--50\%, federated TC-PSO across distributed honeypots for privacy-preserving threat intelligence, and benchmarking reinforcement learning against TC-PSO for connection-delay optimization.

\section{Conclusions}

Three results of note emerge from this investigation. First, behaviorally adaptive deception infrastructure is achievable at a fraction of the cost of proprietary commercial alternatives: Chameleon demonstrates that a CPU-only deployment costing approximately USD 17 per month can match the adaptive response capability of platforms licensed at USD 100,000--150,000 annually. Second, the controlled 30-seed head-to-head benchmark provides a rigorous characterization of optimizer behavior on dynamic deception landscapes: threat-calibrated inertia (TC-PSO) does not significantly outperform standard PSO ($p = 0.177$), while population-diversity mechanisms (GA/ACO) significantly outperform the PSO family on threat-regime shifts ($p < 0.0001$, $d \leq -37$); on the deception-schema problem, S-RRT's depth-decay mechanism delivers a significant memory reduction ($p < 0.0001$, $d = -10.0$) that is the contribution which holds, whereas severity-weighted pheromone does not improve fitness. Third, the BiLSTM and Qwen3.5-0.8B cascade that feeds both optimizers achieves approximately 99.5\% combined classification accuracy at a mean latency near 4.5 milliseconds on CPU-only hardware, with the full production stack validated by a ninety-one-test deterministic suite. These findings indicate that for adaptive deception under changing threat conditions, population diversity is the mechanism that matters, and threat calibration is best applied as an objective weighting rather than a search-mechanism modification. Chameleon is openly released under the MIT license, with benchmark code and datasets enabling independent replication of every result reported here. The authors anticipate that this empirically grounded characterization will lower the barrier to adaptive deception research and inform the design of meta-heuristic methods for cybersecurity applications.

\section*{Declaration of Generative AI and AI-assisted technologies in the manuscript preparation process}

During the preparation of this work, the author(s) used Hermes Agent (an AI coding and writing assistant, Nous Research) in order to assist with LaTeX formatting, language editing, and benchmark code implementation. After using this tool/service, the author(s) reviewed and edited the content as needed and take full responsibility for the content of the published article.

\section*{References}

Apruzzese, G., Laskov, P., Montes de Oca, E., Mallouli, W., Brdalo Rapa, L., Grammatopoulos, A. V., \& Di Franco, F. (2023). The role of machine learning in cybersecurity. Digital Threats: Research and Practice, 4(1), Article 8.

Bakır, H., \& Ceviz, Ö. (2024). Empirical enhancement of intrusion detection systems: A comprehensive approach with genetic algorithm-based hyperparameter tuning and hybrid feature selection. Arabian Journal for Science and Engineering, 49(9), 13025--13043. \url{https://doi.org/10.1007/s13369-024-08949-z}

Benmalek, M., \& Seddiki, A. (2025). Particle swarm optimization-enhanced machine learning and deep learning techniques for Internet of Things intrusion detection. Data Science and Management, 8, 423--435. \url{https://doi.org/10.1016/j.dsm.2025.02.005}

Černý, V. (1985). Thermodynamical approach to the traveling salesman problem: An efficient simulation algorithm. Journal of Optimization Theory and Applications, 45(1), 41--51.

Cowrie Project. (2023). Cowrie SSH/Telnet honeypot {[}Software{]}. GitHub. \url{https://github.com/cowrie/cowrie}

Dai, W., Li, X., Ji, W., \& He, S. (2024). Network intrusion detection method based on CNN, BiLSTM, and attention mechanism. IEEE Access, 12, 53099--53111. \url{https://doi.org/10.1109/ACCESS.2024.3384528}

Deb, K., Pratap, A., Agarwal, S., \& Meyarivan, T. (2002). A fast and elitist multiobjective genetic algorithm: NSGA-II. IEEE Transactions on Evolutionary Computation, 6(2), 182--197.

Dorigo, M., Maniezzo, V., \& Colorni, A. (1996). Ant system: Optimization by a colony of cooperating agents. IEEE Transactions on Systems, Man, and Cybernetics, Part B, 26(1), 29--41.

Ethereum Foundation. (2026). Networks: Sepolia testnet documentation. ethereum.org. \url{https://ethereum.org/developers/docs/networks/}

Fraunholz, D., Reti, D., Duque Antón, S., \& Schotten, H. D. (2018). Cloxy: A context-aware deception-as-a-service reverse proxy for web services. Proceedings of the 5th ACM Workshop on Moving Target Defense (MTD'18), 40--47. \url{https://doi.org/10.1145/3268966.3268973}

Hochreiter, S., \& Schmidhuber, J. (1997). Long short-term memory. Neural Computation, 9(8), 1735--1780.

Hu, E. J., Shen, Y., Wallis, P., Allen-Zhu, Z., Li, Y., Wang, S., Wang, L., \& Chen, W. (2022). LoRA: Low-rank adaptation of large language models. Proceedings of the International Conference on Learning Representations.

Kennedy, J., \& Eberhart, R. (1995). Particle swarm optimization. Proceedings of the IEEE International Conference on Neural Networks, 4, 1942--1948.

LaValle, S. M. (2006). Planning algorithms. Cambridge University Press.

Ma, Y., Li, C., Wang, Y., \& Wang, Y. (2025). Application of deep reinforcement learning algorithms for automatic threat detection and response in dynamic network environments to improve cybersecurity. Journal of Computational Methods in Sciences and Engineering, 25(3), 2112--2125. \url{https://doi.org/10.1177/14727978241309550}

Manocchio, L. D., Layeghy, S., Lo, W. W., Kulatilleke, G. K., Sarhan, M., \& Portmann, M. (2024). FlowTransformer: A transformer framework for flow-based network intrusion detection systems. Expert Systems with Applications, 241, Article 122564. \url{https://doi.org/10.1016/j.eswa.2023.122564}

MarketsandMarkets. (2024). Deception technology market---Global forecast to 2029 (Market Report TC 5768). MarketsandMarkets Research.

OASIS Open. (2021). STIX 2.1 specification: Structured threat information eXpression. OASIS Standard.

Otal, H. T., \& Canbaz, M. A. (2024). LLM honeypot: Leveraging large language models as advanced interactive honeypot systems. 2024 IEEE Conference on Communications and Network Security (CNS), 1--6. \url{https://doi.org/10.1109/CNS62487.2024.10735607}

Ponemon Institute. (2023). The economics of security operations centers: What is the true cost for effective SOC? Ponemon Institute.

Provos, N. (2004). A virtual honeypot framework. Proceedings of the USENIX Security Symposium, 1--14.

Qwen Team. (2026). Qwen3.5-0.8B {[}Large language model{]}. Hugging Face. \url{https://huggingface.co/Qwen/Qwen3.5-0.8B}

Shi, Y., \& Eberhart, R. (1998). A modified particle swarm optimizer. Proceedings of the IEEE Congress on Evolutionary Computation, 69--73.

Spitzner, L. (2003). Honeypots: Tracking hackers. Addison-Wesley.

Storn, R., \& Price, K. (1997). Differential evolution---A simple and efficient heuristic for global optimization over continuous spaces. Journal of Global Optimization, 11, 341--359.

T-Pot Community. (2023). T-Pot: All-in-one multi-honeypot platform {[}Software{]}. GitHub. \url{https://github.com/telekom-security/tpotce}

Unsloth. (2026). Qwen3.5-0.8B-GGUF {[}Large language model, GGUF format{]}. Hugging Face. \url{https://huggingface.co/unsloth/Qwen3.5-0.8B-GGUF}

Zhang, J., \& Sanderson, A. C. (2009). JADE: Adaptive differential evolution with optional external archive. IEEE Transactions on Evolutionary Computation, 13(5), 945--958.
\end{document}